\documentstyle[epsfig,12pt,dina4]{article}

\begin{document}

\baselineskip=18pt	

\begin{center}

{\Large\bf
New Temperature Measurements 
\vspace{0.2cm}

in $^{197}$Au + $^{197}$Au Collisions}

\vspace*{0.6cm}

\normalsize                

W. Trautmann$^{(1)}$ for the ALADIN collaboration:
\vspace*{0.1cm}

R.~Bassini,$^{(2)}$
M.~Begemann-Blaich,$^{(1)}$
A.S.~Botvina,$^{(3)}$\footnote{
Present Address: Bereich Theoretische Physik, Hahn-Meitner-Institut,
D-14109 Berlin, Germany}
S.~Fritz,$^{(1)}$
S.J.~Gaff,$^{(4)}$
C.~Gro\ss,$^{(1)}$
G.~Imm\'{e},$^{(5)}$
I.~Iori,$^{(2)}$
U.~Kleinevo\ss,$^{(1)}$
G.J.~Kunde,$^{(4)}$
W.D.~Kunze,$^{(1)}$
U.~Lynen,$^{(1)}$
V.~Maddalena,$^{(5)}$			
M.~Mahi,$^{(1)}$
T.~M\"ohlenkamp,$^{(6)}$
A.~Moroni,$^{(2)}$
W.F.J.~M\"uller,$^{(1)}$
C.~Nociforo,$^{(5)}$       		
B.~Ocker,$^{(7)}$
T.~Odeh,$^{(1)}$
F.~Petruzzelli,$^{(2)}$
J.~Pochodzalla,$^{(1)}$\footnote{
Present address: Max-Planck-Institut f\"ur Kernphysik,
D-69117 Heidelberg, Germany}
G.~Raciti,$^{(5)}$
G.~Riccobene,$^{(5)}$			
F.P.~Romano,$^{(5)}$    		
Th.~Rubehn,$^{(1)}$\footnote{
Present address: Nuclear Science Division, Lawrence Berkeley Laboratory,
Berkeley, CA 94720, USA}
A.~Saija,$^{(5)}$     			
M.~Schnittker,$^{(1)}$
A.~Sch\"uttauf,$^{(7)}$
C.~Schwarz,$^{(1)}$
W.~Seidel,$^{(6)}$
V.~Serfling,$^{(1)}$
C.~Sfienti,$^{(5)}$			
A.~Trzcinski,$^{(8)}$
G.~Verde,$^{(5)}$
A.~W\"orner,$^{(1)}$
Hongfei~Xi,$^{(1)}$\footnote{
Present address: NSCL, Michigan State University,
East Lansing, MI 48824, USA }
and B.~Zwieglinski$^{(8)}$

\vspace{0.5cm}

$^{(1)}$Gesellschaft  f\"ur  Schwerionenforschung, D-64291 Darmstadt, 
Germany\\
$^{(2)}$Istituto di Scienze Fisiche dell' Universit\`{a} 
and I.N.F.N., I-20133 Milano, Italy\\
$^{(3)}$Institute for Nuclear Research,
Russian Academy of Sciences, 117312 Moscow , Russia\\
$^{(4)}$Department of Physics and
Astronomy and National Superconducting Cyclotron Laboratory,
Michigan State University, East Lansing, MI 48824, USA\\
$^{(5)}$Dipartimento di Fisica dell' Universit\`{a}
and I.N.F.N.,
I-95129 Catania, Italy\\
$^{(6)}$Forschungszentrum Rossendorf, D-01314 Dresden, Germany\\
$^{(7)}$Institut f\"ur Kernphysik,
Universit\"at Frankfurt, D-60486 Frankfurt, Germany\\
$^{(8)}$Soltan Institute for Nuclear Studies,
00-681 Warsaw, Hoza 69, Poland

\vspace{0.6cm}

{ABSTRACT}
\end{center}

\normalsize                

We report on new measurements of 
breakup temperatures for target spectators from 
$^{197}$Au~+~$^{197}$Au reactions at 1000 MeV per nucleon. 
The temperatures rise with decreasing impact parameter from
4 MeV for peripheral to about 10 MeV for the most central collisions,
in good agreement with previous results for
projectile spectators at 600 MeV per nucleon.

The measured temperatures agree quantitatively with the breakup temperatures
predicted by the statistical multifragmentation model.
For these calculations a relation between the initial 
excitation energy and mass 
was derived which gives good simultaneous agreement 
for the fragment charge correlations.

The energy spectra of light charged particles and 
the behaviour of the mean kinetic energies of neutrons
indicate a substantial component of light particle emission 
prior to the final breakup stage.

\newpage

\section{Introduction}
\label{Sec_1}

The correlated measurements of the nuclear temperature and the excitation 
energy for excited projectile spectators 
in $^{197}$Au + $^{197}$Au collisions 
at 600 MeV per nucleon has permitted to extend the caloric curve of 
nuclei to very high excitation energies, far beyond the nuclear binding
energy \cite{poch95,theo95}. 
For the temperature determination the method suggested
by Albergo {\it et al.} has been
used which is based on the assumption of chemical equilibrium and
requires the measurement of double ratios of 
isotopic yields \cite{albergo}. The obtained
temperatures, plotted against the measured excitation energy, resulted in a
caloric curve with the characteristic behavior reminiscent of 
first-order phase transitions in macroscopic systems.
The 'liquid' and the 'vapor' regimes where
the temperature rises with increasing excitation energy are separated
by a region of nearly constant temperature $T \approx$ 5 MeV over which
the multiplicity of the fragmentation products increases. 

The present discussion of the caloric curve mainly centers around three
points. The first one concerns the methodical question of whether 
breakup temperatures can be reliably deduced from
isotopic yield ratios in the presence of sequential-decay processes
by which these ratios may be modified after the breakup has occurred
\cite{gulm97,tsang97}. Model calculations show that the amount of 
modification depends on the breakup density 
which is not very well known [2,6-8].
The potential directions of interpretation, in a wider sense, 
of these new experimental results
constitute the second topic of discussion [9-11].
Finally, there is numerous
work concerned with the consequences for finite 
nuclei \cite{gross90,bond95} of the predicted first-order phase 
transition in nuclear matter \cite{jaqaman,brack} 
(for very recent references see, e.g., refs. \cite{lee97,de97}). In addition,
experimental caloric curves for other systems have been reported
recently \cite{haug96,ma97}.

In this contribution, we present results of new measurements of the breakup
temperature in the $^{197}$Au + $^{197}$Au reaction. 
In particular, we will address two specific questions connected
with the statistical interpretation of multi-fragment decays of excited
spectator systems: Do the measured temperatures exhibit the same
invariance with respect to the entrance channel that has been
found for the fragmentation patterns? This universal feature of the 
spectator decay has become evident in the observed $Z_{bound}$ scaling
of the fragment multiplicities and charge correlations \cite{schuetti}.
Here $Z_{bound}$, defined as the sum of the atomic numbers
$Z_i$ of all spectator fragments with $Z_i \geq$ 2, is a quantity
closely correlated with the excitation energy transferred during the
initial stages of the reaction. Secondly, how do the extracted
temperatures compare with the predictions of the statistical 
multifragmentation model \cite{bond95} which has been so successfull
in describing the multiplicities and charge correlations 
characterizing the universal partition space \cite{botv95}?

\section{Experimental method and results}
\label{Sec_2}

The experiment was performed at the ALADIN spectrometer \cite{schuetti} 
of the GSI facility with beams of $^{197}$Au of 1000 MeV per
nucleon incident energy, provided by the heavy-ion
synchrotron SIS.
The present data were taken as part of a larger
experiment which incorporated three multi-detector 
hodoscopes for coincident 
particle detection \cite{fritz97,schwarz97}.

A set of seven telescopes, each consisting of three Si detectors with 
thickness 50, 300, and 1000 $\mu$m and of a 4-cm long CsI(Tl) scintillator
with photodiode readout, were used to
measure the isotopic yields of light charged particles and fragments. 
Four telescopes were placed in the forward hemisphere
while three telescopes
were placed at $\theta_{lab}$ = 110$^{\circ}$, 130$^{\circ}$, 
and 150$^{\circ}$ for detecting the products of the target-spectator
decay. 
Isotopes in the range from hydrogen to carbon were
satisfactorily resolved. Because 
of the presumably different shapes of
$^3$He and $^4$He spectra at low energies,
a correction was required in order to compensate for the effect of the
detection threshold $E \approx$ 12 MeV for helium ions, 
resulting from triggering with the
300-$\mu$m detector. 
For a global characterization of the reaction and
impact parameter selection,
the quantity $Z_{bound}$ of the coincident projectile decay was measured
with the time-of-flight (TOF) wall of the ALADIN spectrometer. 

Emission temperatures $T$ were deduced from the double ratios $R$ formed
by combining the ratio of $^{3}$He/$^{4}$He yields with
either the lithium yield ratio 
$^{6}$Li/$^{7}$Li or with the hydrogen yield ratios p/d or d/t. The 
set of $^{3}$He, $^{4}$He, $^{6}$Li, and $^{7}$Li isotopes is the one
used previously for the determination of breakup temperatures of
projectile spectators in $^{197}$Au + $^{197}$Au at 600 MeV 
per nucleon \cite{poch95,theo95}.
Combinations involving p, d, or t, together with $^{3}$He and $^{4}$He,
have the advantage of larger production cross sections, particularly 
in the 'vapor' regime where the 
heavier fragments are becoming rare. 
The three expressions are
\begin{equation}
T_{{\rm HeLi},0}= 13.3/ \ln(2.2\frac{Y_{^{6}{\rm Li}}/Y_{^{7}{\rm Li}}}
{Y_{^{3}{\rm He}}/Y_{^{4}{\rm He}}})
\label{EQ1}
\end{equation}
and
\begin{equation}
T_{{\rm Hepd},0}= 18.4/ \ln(5.6\frac{Y_{^{1}{\rm H}}/Y_{^{2}{\rm H}}}
{Y_{^{3}{\rm He}}/Y_{^{4}{\rm He}}})
\label{EQ2}
\end{equation}
and
\begin{equation}
T_{{\rm Hedt},0}= 14.3/ \ln(1.6\frac{Y_{^{2}{\rm H}}/Y_{^{3}{\rm H}}}
{Y_{^{3}{\rm He}}/Y_{^{4}{\rm He}}})
\label{EQ3}
\end{equation}
where the temperatures are given in units of MeV.

\begin{figure}[ttb]
   \centerline{\epsfig{file=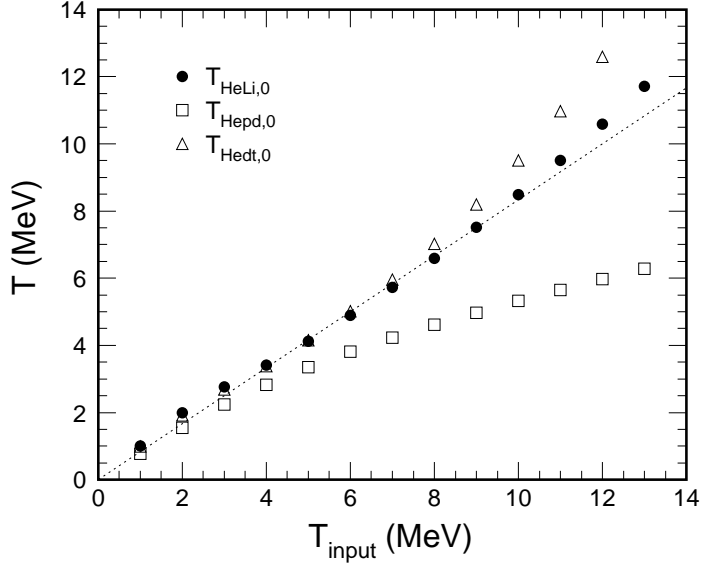,height=8cm}}
        \caption[]{\it\small
Apparent temperatures $T_{{\rm HeLi},0}$, $T_{{\rm Hepd},0}$,
and $T_{{\rm Hedt},0}$,
according to the quantum statistical model,
as a function of the input temperature $T_{input}$. A breakup
density $\rho/\rho_0$ = 0.3 is assumed. The dotted line represents the 
linear
relation $T_0 = T_{input}$/1.2.
        }
        \label{fig1}
\end{figure}

The subscript 0 is meant to indicate that these apparent temperatures,
derived from the measured ground-state populations, may be affected by
feeding of these populations from sequentially decaying 
excited states. 
The required corrections were calculated 
with the quantum statistical model which starts from
chemical equilibrium at a given temperature, density, and neutron-to-proton 
(N/Z) ratio and which includes sequential decay \cite{hahn88,konop94}.
In Fig. 1, the three apparent temperatures defined in eqs. (1-3) 
are shown as a function of the 
equilibrium temperature $T_{input}$ for the parameters $N/Z$ = 1.49 
(value of $^{197}$Au) and density
$\rho = 0.3 \cdot \rho_0$ (where $\rho_0$ is the saturation
density of nuclei). 
The relations between $T_{{\rm HeLi},0}$ or $T_{{\rm Hedt},0}$ and 
$T_{input}$ are almost linear and the corrections required
in these two cases are practically identical, except at the highest
temperatures. The linear approximation, indicated by the dotted line,
corresponds to the constant correction factor $T = 1.2\cdot T_{0}$ 
adopted for $T_{{\rm HeLi}}$ in Ref. \cite{poch95}.
Apparently, $T_{{\rm Hepd},0}$ is
more strongly affected by feeding effects 
at the higher temperatures.
Within the range of densities 0.1 $\le \rho /\rho_0 \le$ 0.5,
the corrections required according to the quantum statistical model 
vary within about $\pm$15\% \cite{theo95,theo}.
They virtually do not change with
the $N/Z$ ratio of the primary source.
In the analysis, the corrections calculated for 
$\rho /\rho_0$ = 0.3, as derived from the results 
shown in Fig. 1, were applied.

\begin{figure}[ttb]
   \centerline{\epsfig{file=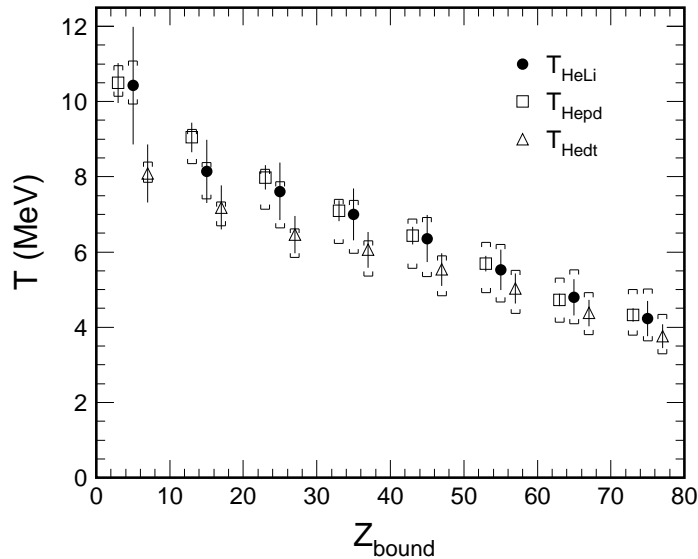,height=8cm}}
        \caption[]{\it\small
Temperatures $T_{{\rm HeLi}}$, $T_{{\rm Hepd}}$, and $T_{{\rm Hedt}}$
as a function of $Z_{bound}$, averaged over bins of 10-units width.
Corrections have been applied as described in the text.
The error bars represent
the statistical uncertainty. The systematic uncertainty, caused by the
extrapolation of the yields of helium isotopes below the identification
threshold, is indicated by the brackets.
For clarity, the open data symbols are laterally displaced by 2 units
of $Z_{bound}$.
        }
        \label{fig2}
\end{figure}

In Fig. 2 the obtained temperatures $T_{{\rm HeLi}}$, 
$T_{{\rm Hepd}}$, and $T_{{\rm Hedt}}$ 
are shown as a function of $Z_{bound}$. They are based on the yield ratios
measured with the telescope at the most backward angle 
$\theta_{lab}$ = 150$^{\circ}$.
Simulations indicate that, at this angle, contributions from the
midrapidity source are negligible.
The temperatures increase continuously with decreasing $Z_{bound}$
from $T$ = 4 MeV for peripheral collisions to about 10 MeV
for the most central collisions associated with the smallest
$Z_{bound}$ values. The range $Z_{bound}\le$ 20 corresponds to
the high excitation energies at which the upbend of the temperature appears 
in the caloric curve \cite{poch95}. The 
results obtained with the three different double ratios agree rather well
which is remarkable in view of the different feeding corrections 
that are required, in particular, in the case of $T_{{\rm Hepd}}$.

\section{Universality of spectator temperatures}
\label{Sec_3}

The universality of spectator fragmentation has first been recognized
in the study of the charge partitions \cite{schuetti,traut95}. 
Universality, in the present 
context, refers to the invariance of the fragmentation patterns of
excited spectator nuclei with respect to the properties of the entrance 
channel. The decaying system has lost the memory of how it was formed.
This was shown to be valid for the
variation of the bombarding energy (beyond about 400 MeV per nucleon)
and for the variation of the target mass (if we consider the case of 
projectile fragmentation). If a linear scaling law is applied, 
it is even valid for the variation of the projectile mass itself.
The search for the deeper symmetries behind these universal properties
of the fragment decay is presently pursued by several groups
\cite{botet93,rich97}.

\begin{figure}[ttb]
   \centerline{\epsfig{file=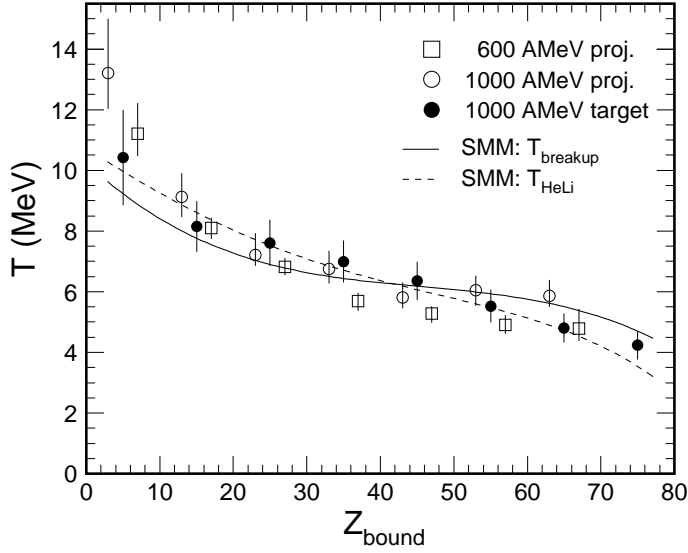,height=8cm}}
        \caption[]{\it\small
Temperatures $T_{{\rm HeLi}}$ of the target spectator from the
present experiment at $E/A$ = 1000 MeV (dots)
and of the projectile spectator (from Ref. \cite{theo})
at $E/A$ = 600 MeV (open squares) and 1000 MeV (open circles)
as a function of $Z_{bound}$. The data symbols represent averages over
bins of 10-units width.
For clarity, the open data symbols are laterally displaced by 2 units
of $Z_{bound}$.
The full and dashed lines
represent the breakup temperature $T_{breakup}$ and the
isotopic temperature $T_{{\rm HeLi}}$ calculated with the
statistical multifragmentation model.
        }
        \label{fig3}
\end{figure}

The data show that the invariance with respect to the entrance
channel includes the deduced isotope temperatures.
In Fig. 3 the $T_{{\rm HeLi}}$ temperatures from this work
and those derived previously \cite{poch95,theo,theo1}
for projectile spectators in $^{197}$Au + $^{197}$Au collisions at
600 and 1000 MeV per nucleon are shown in comparison. In the case
of the projectile decays,
the isotopes were identified by tracking of their
trajectories with the upgraded TP-MUSIC detector and subsequent momentum 
and time-of-flight analysis.
The agreement obtained in the measurements at 1000 MeV per nucleon for
the projectile and the target decays, performed with different 
methods of isotope identification and associated with different 
detection thresholds, reflects the experimental accuracy.
The agreement of the data measured at 600 and 1000 MeV per nucleon 
confirms the expectation that the breakup temperature, 
as a function of $Z_{bound}$, does not change with the bombarding energy.

\begin{figure}[ttb]
   \centerline{\epsfig{file=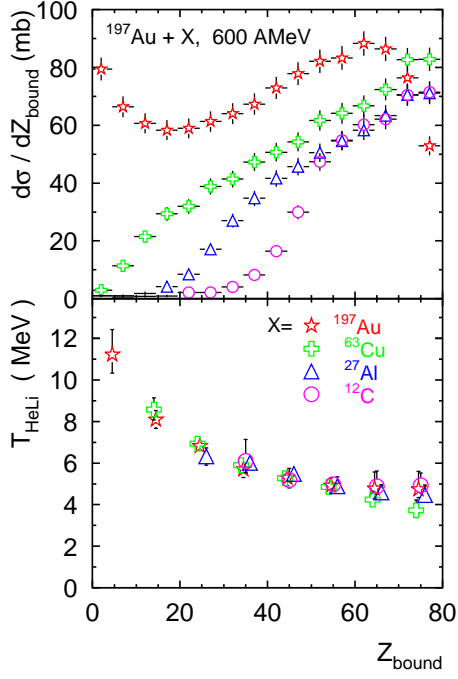,height=9cm}}
        \caption[]{\it\small
Top:
Measured cross sections $d\sigma /dZ_{bound}$
for the reactions of $^{197}$Au projectiles
at $E/A$ = 600 MeV with the four targets
C, Al, Cu, and Au.
Note that the experimental trigger affected the cross sections
for $Z_{bound} \ge$ 65.

\noindent
Bottom:
Breakup temperature $T_{{\rm HeLi}}$ 
for the same four reactions
(from Ref. \cite{theo}).
        }
        \label{fig4}
\end{figure}

Neither does the breakup temperature 
depend on the mass of the collision partner. In Fig. 4 this is 
demonstrated for the case of $^{197}$Au projectiles of 600 MeV per nucleon
impinging on C, Al, Cu, Au targets.
The breakup temperatures deduced by the EOS collaboration
for $^{197}$Au + C at 1 GeV per nucleon \cite{haug96}
are also consistent with this conclusion.
Within the range $Z_{bound} \ge$ 40 which is mainly populated in this
reaction (Fig. 4, top, and Ref. \cite{schuetti}), 
they are in good agreement with the present
results for $^{197}$Au + $^{197}$Au, both in absolute
magnitude and in their dependence on the impact parameter.
The observed universality is a strong indication
for equilibrium being reached at the breakup stage which is a
necessary condition for a statistical description of the fragmentation
process.

\section{Statistical model description}
\label{Sec_4}

Calculations within the statistical multifragmentation model
\cite{bond95} were performed in order to test its consistency with
respect to the statistical parameters and predicted charge partitions.
In the model one assumes that all observed particles come from the decay of one
equilibrated source.

\begin{figure}[ttb]
   \centerline{\epsfig{file=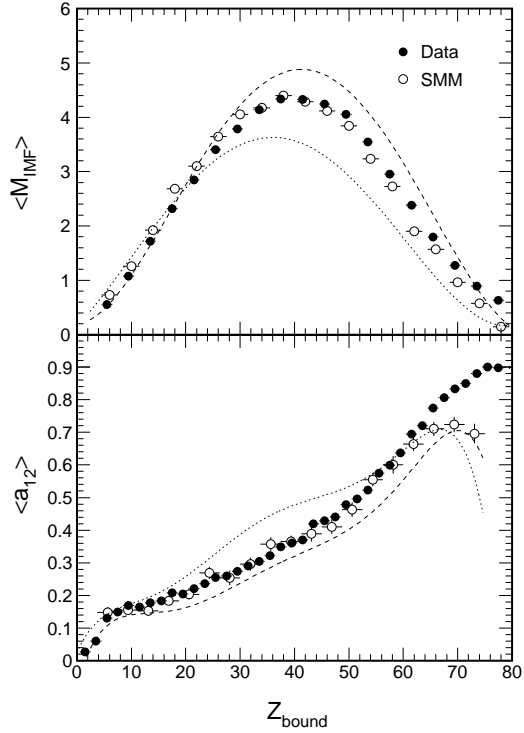,height=10cm}}
        \caption[]{\it\small
Mean multiplicity of intermediate-mass
fragments $\langle M_{IMF} \rangle$ (top)
and mean charge asymmetry $\langle a_{12}\rangle$ (bottom)
as a function of $Z_{bound}$, as obtained from the calculations with the
statistical multifragmentation model (open circles) in comparison to the
experiment (dots, from Ref. \cite{schuetti}). The dashed and dotted lines
show the results of the calculations with excitation energies
$E_x/A$ 15\% above and 15\% below the adopted values, respectively.
Note that the trigger threshold affected the data of Ref. \cite{schuetti}
at $Z_{bound} \ge$ 65.
        }
        \label{fig5}
\end{figure}

As a criterion for selecting the distribution in 
excitation energy and mass of the ensemble of 
excited spectator nuclei, required as input for the calculations,
we chose the capability of the model to simultaneously
describe the correlations of the mean multiplicity 
$\langle M_{IMF}\rangle$ of intermediate-mass fragments (IMF's) 
and of the mean charge asymmetry $\langle a_{12}\rangle$ with 
$Z_{bound}$. The asymmetry $a_{12}$ of the two largest fragments
is defined as $a_{12} = (Z_{max}-Z_2)/(Z_{max}+Z_2)$,
with the mean value
to be calculated from all events with $Z_{max}\ge Z_2 \ge$ 2.
The comparison, shown in Fig. 5, was based on the data reported in 
Ref. \cite{schuetti}.
In the region $Z_{bound}~>$~30
the mean excitation energy of the ensemble of 
spectator nuclei was found to be well
constrained by the mean fragment multiplicity alone.
At $Z_{bound} \approx$ 30 and below, the charge asymmetry 
was a necessary second constraint (cf. Ref. \cite{deses96}) while, at
the lowest values of $Z_{bound}$, neither the multiplicity nor
the asymmetry provided rigid constraints on the excitation energy.
These sensitivities are illustrated in Fig. 5 where the dashed and
dotted lines show the
model results for $E_x/A$ chosen 15\% above and below the adopted values.

The solid line in Fig. 3 represents the thermodynamical 
temperature $T$ obtained from the calculations.
Within the given experimental and methodical uncertainties, it is in
very good agreement with the data. The description
of the fragmentation as a statistical process is thus internally 
consistent, the temperatures needed to reproduce the 
observed partition patterns correspond to those measured.
The comparison is even more consistent if we take directly the
temperature $T_{{\rm HeLi}}$ 
that is obtained from the calculated isotope yields
and corrected in the same way as the experimental data (dashed line).
The agreement of the measured and model $T_{{\rm HeLi}}$ is excellent, 
and the tendency of being lower (higher) 
than the thermodynamical temperature at the larger (smaller) 
values of $Z_{bound}$ is common to both. According to the statistical
multifragmentation model, therefore, the true source temperature
varies only from $T$ = 5 MeV for peripheral to about 9 MeV for the
most central collisions and stays rather constant $T \approx$ 6 MeV 
over a wide range of
$Z_{bound}$, i.e. over a wide range of excitation energies, where the
maximum production of intermediate-mass fragments is observed.

The excitation energies that have resulted from this procedure are
somewhat larger than what was found previously in analyses [21,30-32]
of the earlier $^{197}$Au on Cu data at 600~MeV per nucleon
\cite{kreutz}. The difference reflects the sensitivity
to the fragment multiplicity and is caused by the slightly larger
mean multiplicities that were obtained from 
the more recent experiments with improved acceptance \cite{schuetti}.
They are still considerably lower than the energies obtained from
the calorimetric measurement of the total energy deposit (see section 6).
A (partial) resolution of this discrepancy may come from pre-breakup
emission of light particles carrying away energy
as the spectator system approaches the final breakup stage.

\section{Indications of pre-breakup emission}
\label{Sec_5}

Kinetic energy spectra were studied for light charged particles up 
to $^{4}$He. 
For the five species proton, deuteron, triton, 
$^{3}$He, and $^{4}$He, the spectra measured at $\theta_{lab}$~=~150$^{\circ}$
and integrated over finite regions of
$Z_{bound}$ are shown in Fig. 6. 
For comparison, the predictions calculated with the statistical 
fragmentation model are also given.
The two sets of experimental and model spectra are each normalized,
and one overall normalization factor
is used to relate the two sets.
It was adapted to the yields of $Z$ = 2 fragments
because their calculated multiplicities,
as a function of $Z_{bound}$, are found to satisfactorily
reproduce the experimental multiplicities reported in Ref. \cite{schuetti}.

\begin{figure}[ttb]
   \centerline{\epsfig{file=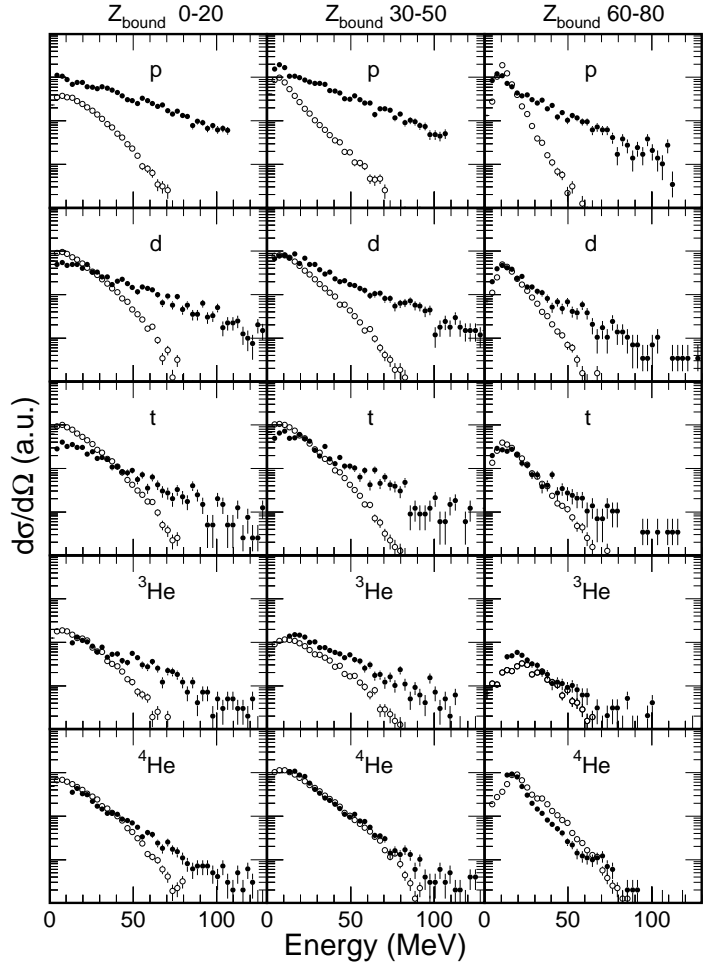,height=13cm}}
        \caption[]{\it\small
Energy spectra, measured at $\theta_{lab} = 150^{\circ}$,
of light charged particles p, d, t, $^3$He,
and $^4$He for three
intervals of $Z_{bound}$ as indicated. The dots represent the
measured spectra, the open circles are the results of the calculations
with the statistical multifragmentation model. The spectra are
normalized as stated in the text.
        }
        \label{fig6}
\end{figure}

The main trend apparent from the comparison is a systematically
increasing deviation of the experimental from the model spectra
with decreasing $Z_{bound}$, i.e. increasing centrality, and
with decreasing particle mass. It not only affects the slope parameters
describing the shape of the spectra but also the integrated intensities.
The yields of hydrogen isotopes, and in particular of the protons, are
grossly underestimated by the statistical multifragmentation model.
In the case of $^4$He, on the other hand, 
the equilibrium description accounts rather well for the 
multiplicities and kinetic energies.
A major contribution to the observed $^4$He yields is expected to come
from evaporation by large fragments and excited residue-like 
nuclei which, apparently, is modelled well. 

Conceivable mechanisms that cannot explain the observed deviations
include collective flow and Coulomb effects which both should act
in proportion to the mass or charge of the emitted particle, 
contrary to what is observed. 
On the other hand, the commonly adopted scenario of
freeze-out after expansion involves a pre-breakup phase during which the
system cools not only by adiabatic expansion but also by the emission
of light particles, predominantly nucleons but also light complex
particles \cite{papp95,botv92,fried90}.
The spectra should reflect the
higher temperatures at the earlier stages of the reaction,
prior to the final breakup into fragments, but not necessarily exhibit
a clear preequilibrium character (cf. Ref. \cite{kwiat97}).

\begin{figure}[ttb]
   \centerline{\epsfig{file=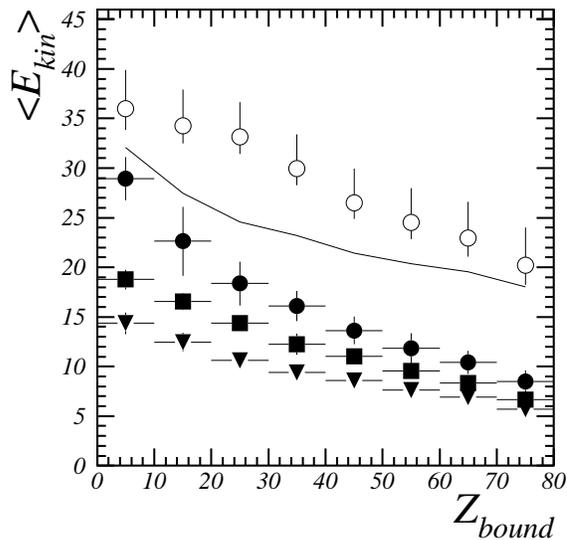,height=7.5cm}}
        \caption[]{\it\small
Mean kinetic energy of neutrons (full symbols) 
in the rest frame of the projectile
spectator as obtained in measurements with LAND for the three
bombarding energies 600 (triangles), 800 (squares), and 1000 MeV per 
nucleon (circles).
For the case of 1000 MeV per nucleon, a comparison is made between the
measured mean kinetic energies of protons (open circles)
and those obtained by adding estimated Coulomb contributions to the
neutron kinetic energies (full line).
        }
        \label{fig7}
\end{figure}

A significant component of pre-breakup emission in the light particle
yields has two consequences that deserve particular 
attention. The pre-breakup yields of protons, deuterons, and tritons
are included in the double ratios used to determine the temperatures
$T_{{\rm Hepd}}$ and $T_{{\rm Hedt}}$. This violates the requirement of
thermal and chemical equilibrium, which is the basic assumption
of the method, and thus may shed doubt on the meaning of the consistency 
exhibited by Fig. 2. On the other hand, the
quantum statistical model predicts that, in particular, 
the p/d ratio varies sufficiently slowly with temperature,
such that the overall ratio will not be significantly affected by 
contributions from higher temperatures. 
The second point concerns the interpretation of the excitation energy
that is obtained in a calorimetric measurement.

\section{Excitation energy of primary spectators}
\label{Sec_6}

Rather small fractions of the initial bombarding energy are imparted to the
spectator nuclei in relativistic collisions. The actual amount
of energy deposition
can only be reconstructed from the exit-channel configuration which
requires a complete knowledge of all decay products, including their
atomic numbers, masses, and kinetic energies. A rather large fraction
of it resides in the kinetic energies of the produced nucleons and in their
separation energies.

\begin{figure}[ttb]
   \centerline{\epsfig{file=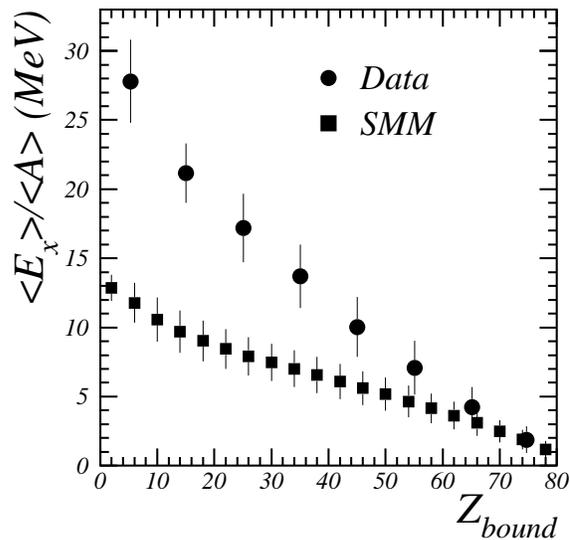,height=7.5cm}}
        \caption[]{\it\small
Reconstructed average excitation energy
$\langle E_x \rangle / \langle A \rangle$ of the decaying
spectator system (circles) and mean
excitation energy used in the calculations with the statisistical
multifragmentation model as a function of $Z_{bound}$.
        }
        \label{fig8}
\end{figure}

For the analysis for the $^{197}$Au + $^{197}$Au reaction
at 600 MeV per nucleon, the data on neutron
production measured with LAND were available \cite{poch95,traut95}.
On the other hand, the hydrogen isotopes were not detected in this
experiment and assumptions concerning their kinetic energies as well
as the p/d/t intensity ratios and the overall $N/Z$ ratio of 
the spectator had to be made.
From the present experiment at 1000 MeV per nucleon, 
data on hydrogen emission from the
target spectator source have been obtained. Together with the neutron
data measured with LAND at the same bombarding energy in the 
previous experiment, a complete picture on light particle emission has 
become available for this reaction. It was found, in particular, that 
by adding a Coulomb component to the neutron kinetic energies,
an assumption that had to be made previously, the proton kinetic 
energies are considerably underestimated (Fig. 7).
Furthermore, the full analysis of neutron emission at three bombarding
energies has revealed a significant dependence of the neutron 
kinetic energies on the bombarding energy. This phenomenon contrasts
the universal properties of fragment production \cite{schuetti} but
is clearly consistent with a pre-breakup component of nucleon emission
exhibiting some residual memory of the initial stages of the reaction.

In Fig. 8 the result of the present analysis of the spectator energy
and the excitation energies used for the calculations
with the statistical multifragmentation model are shown. In both cases, the
values have increased considerably, compared to those reported
previously \cite{schuetti}. The reasons have been given already. 
The masses $A_0$ of the primary spectator systems, on the other hand, 
have remained the same 
as they consist mainly of the sum of the masses of the produced fragments
of intermediate mass.
They are in rather good agreement with the predictions of the geometrical
participant-spectator model \cite{gosset}.
The maximum number of fragments, observed at $Z_{bound} \approx$ 40, is
associated with initial excitation energies of
$\langle E_x \rangle/\langle A \rangle \approx$ 12 MeV and,
according to the model calculations, with excitation energies of
$\langle E_x \rangle/\langle A \rangle \approx$ 6 MeV in the equilibrium
stage at breakup.

\section{Summary}
\label{Sec_7}

Breakup temperatures $T_{{\rm HeLi}}$, $T_{{\rm Hepd}}$, 
and $T_{{\rm Hedt}}$ were measured for target 
spectators in $^{197}$Au~+~$^{197}$Au collisions at 1000 MeV per nucleon.
In these reactions multifragmentation is the dominant decay channel of 
the produced spectator systems over a wide range of
excitation energy and mass. The corrections for
sequential feeding of the ground-state yields, based on calculations
with the quantum statistical model, resulted in mutually consistent
values for the three temperature observables.

With decreasing $Z_{bound}$, the obtained temperatures increase 
from $T$ = 4 MeV for peripheral collisions to about 10 MeV
for the most central collisions.
Within the errors,
these values are in good agreement with those measured 
previously with the ALADIN spectrometer for projectile spectators
in the same reaction at 600 and 1000 MeV per nucleon.
The agreement of the temperatures measured at 600 and 1000 MeV per nucleon
confirms the expected invariance of the breakup temperature
with the bombarding energy. 
It is consistent with 
the observed $Z_{bound}$ scaling
of the mean fragment multiplicities and charge correlations 
and supports the statistical interpretation of the multi-fragment 
decay of highly excited spectator nuclei.

The comparison with the results of calculations within the
statistical multifragmentation model shows that a good simultaneous 
agreement for the charge partitions and for the breakup
temperatures can be achieved. A necessary requirement
for a consistent statistical description of the spectator fragmentation
is thus fulfilled. 
The model results also suggest that 
the true source temperature varies slightly less, from $T$ = 5 MeV 
for peripheral to only about 9 MeV for the
most central collisions, and stays rather constant $T \approx$ 6 MeV 
over the range of excitation energies where the
maximum production of intermediate-mass fragments is observed.

The systematic behavior of the kinetic-energy spectra of light
charged particles indicates contributions 
from light-particle emission prior to the final
breakup stage. This is supported by the variation with bombarding energy 
of the kinetic-energy spectra of neutrons. A more quantitative
understanding of the role of the pre-breakup processes will be 
essential for the interpretation of temperatures obtained from 
light-particle yields as well as of the excitation energies
obtained from calorimetric measurements of the spectator source. 

\vspace{0.2cm}

\renewcommand{\baselinestretch}{0.85}
\Large                
\normalsize                


\begin{thebibliography}{99}
\bibitem{theo1}
The projectile temperatures given in Fig. 3 are from 
Ref. \cite{theo}. At small
$Z_{bound}$, i.e. at large excitation energies, they exceed those
presented in Ref. \cite{poch95} by 10 to 20\%.
\end{thebibliography}
\end{document}